\documentclass[11pt]{article}
\usepackage{epsfig}
\usepackage{epsf}
\usepackage{multicol}

\long\def\symbolfootnote[#1]#2{\begingroup%
\def\thefootnote{\fnsymbol{footnote}}\footnote[#1]{#2}\endgroup}

\newcommand {\cxpp}[1]{\chi_{c#1}\to \phi\phi}

\newcommand { \tpz} {^3P_0}

\newcommand { \ccbar} {c\bar c}

\newcommand {\sla}[1]{ #1 \!\!\!/}

\newcommand { \ccj } {\chi_{cJ}}
\newcommand {\chicj}{\ccj}

\newcommand { \ccjpp} {\ccj \to \phi\phi}

\begin{document}

\begin{center}
{\Large \bf Mechanisms for $\chi_{cJ}\rightarrow \phi\phi$ Decays} \\
\vspace{1cm} H. Q. Zhou$^{2}$\symbolfootnote[2]{E-mail:
zhouhq@mail.ihep.ac.cn},R. G. Ping$^{1,2}$,
B. S. Zou$^{1,2}$ \symbolfootnote[3]{E-mail: zoubs@mail.ihep.ac.cn}\\
1)CCAST (World Lab.), P.O.Box 8730, Beijing 100080, P.R.China \\
2)Institute of High Energy Physics, P.O.Box 918(4), Beijing
100049,P.R.China
\vspace{1cm}\\
\end{center}

\begin{abstract}
Exclusive decays of $\chi_{cJ}(J=0,2)$ into $\phi\phi$  are
investigated in the framework of perturbative quantum
chromodynamics(pQCD) and $\tpz$ quark pair creation model. The
results show that these two mechanisms exhibit a quite different
behavior in evaluating the decay width for the $\chi_{c0}$ and
$\chi_{c2}$. In pQCD method with nonrelativistic(NR)
approximation, while the calculated $\cxpp{2}$ decay width is
comparable with measured one, the result for the $\cxpp{0}$ decay
width is suppressed and much smaller than experimental value.
However, in $\tpz$ quark pair creation model, the situation is
reversed: the decay width of  $\cxpp{0}$ is greatly enhanced and
can reproduce the large measured value, while the contribution to
the $\cxpp{2}$ decay width is small. The results suggest that
while the pQCD mechanism is the dominant mechanism for $\cxpp{2}$
decay, the $\tpz$ quark pair creation
mechanism is the dominant one for $\cxpp{0}$ decay.
\end{abstract}

\vspace{0.5cm}

\newpage

\section{Introduction}
Exclusive decays of charmonium provide useful information on
quark-gluon dynamics and have been a subject attracting people's
interests for many years\cite{kopke,Novirov}. About twenty years
ago, Brodsky and Lepage extensively investigated the exclusive
decays of heavy quarkonium in the framework of perturbative QCD
\cite{Lepage&B}. They argued that the annihilation of $\ccbar$
quarks is a short distant process and can be described by
perturbative QCD because of the large scale of transferred
momentum $Q^2$ involved. The bound properties of hadrons are
parameterized into wave functions. In this theoretical context,
charmonium decays were extensive studied by many authors. For
example, the exclusive decays $J/\psi\to B\bar B$ ($B:$ octet
baryon) were studied with results consistent with the experimental
data\cite{kroll,Pingrg}. However, for exclusive decays of
$\chi_{cJ}~(J=0,2)$ to two pseudo-scalars, many calculations
\cite{kroll,duncan,Anselmino1,huangtao} found that the obtained
results are much smaller than the data. Recently, the color-octet
contributions to the P-wave charmonium decays have received a
renewed interest\cite{Fock}. In this point of view, the
color-singlet and color-octet contributions to $\ccj$ decays are
at the same order, and should be considered for the calculation of
P-wave charmonium decays. Including the color octet contributions,
the exclusive decays $\chi_{cJ}\to\pi\pi,KK,\eta\eta$ $(J=0,2)$
were studied by Bolz et. al \cite{kroll}. While their calculated
results reproduce the experimental values for $\chi_{c2}$ decays,
those for $\chi_{c0}$ decays are still smaller than the values
from PDG. In these calculations, light-front wave-functions are
used for light pseudo-scalar mesons. Other calculation
\cite{Chernyak} indicates that the results may depend on treatment
of these light-front wave functions.

\vspace{0.5cm}
\begin{figure}[htbp]
\parbox{.4\textwidth}{\epsfysize=3.2cm \epsffile{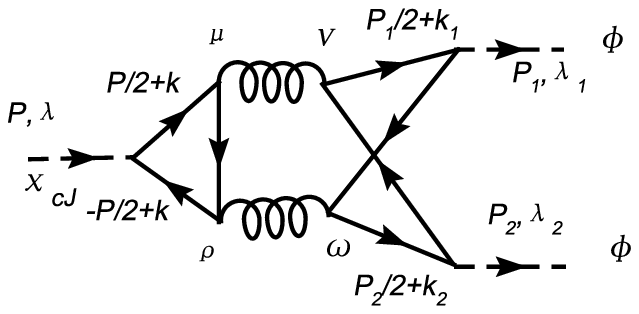}}
\parbox{.4\textwidth}{\epsfysize=3.2cm \epsffile{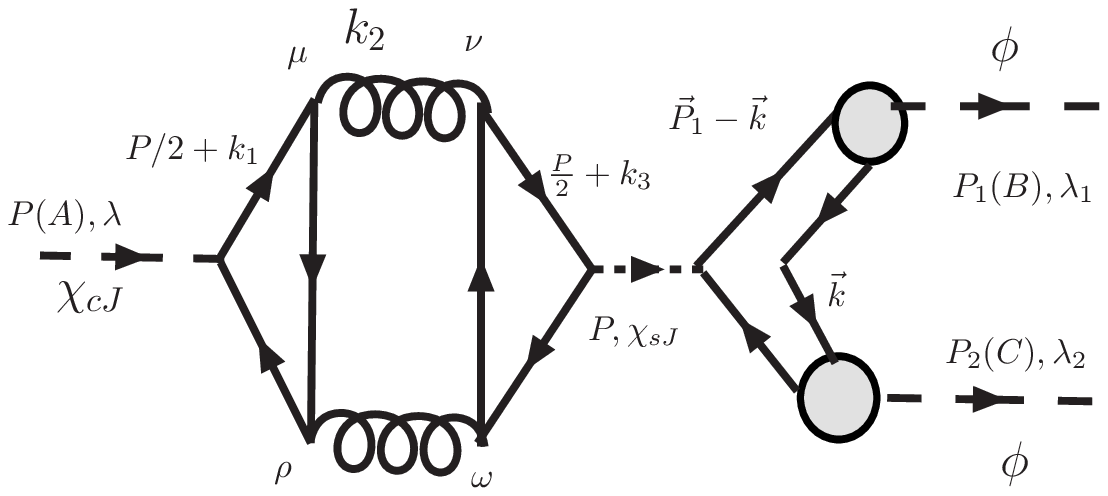}}
\vspace{0.5cm}
\centering{(a)~~~~~~~~~~~~~~~~~~~~~~~~~~~~~~~~~~~~~~~~~~~~~~~~~~(b)}

\caption{\label{fig1} Mechanisms for $\chicj\to\phi\phi$ : (a)
perturbative QCD scheme  and (b) $\tpz$ quark pair creation
scheme.}
\end{figure}

In this work, first we study the conventional pQCD mechanism as
shown in Fig.\ref{fig1}(a) for $\chi_{cJ}\to\phi\phi$. Here for
$\phi$ meson, non-relativistic quark model wave function is
expected to be a reasonable approximation and is used instead of
the light-front wave-function. We meet the same problem as
previous calculations for pseudo-scalar mesons : the result for
$\chi_{c0}$ is too small. This suggests that there should be some
new mechanism playing a role here. So we study a new mechanism as
shown in Fig.\ref{fig1}(b) for the $\chi_{cJ}\to\phi\phi$ : the
$\chi_{cJ}$ first transits to an intermediate $s\bar s$ state
through two gluons, then the intermediate $s\bar s$ state decays
into two $\phi$ mesons by the $\tpz$ quark pair creation mechanism
commonly used in light hadron decays \cite{Ackleh,Capstick}. We
find that this mechanism gives a very large contribution to the
$\chi_{c0}$ decay while it only gives very small contribution to
the $\chi_{c2}$ decay.

The organization of this paper is as follows. In Sec.2 we give the
formulae for calculating the decay width
$\Gamma(\chi_{c0,2}\rightarrow\phi\phi)$ in perturbative QCD
scheme with leading order and NR approximation; in Sec.3 we
develop the formalism for the $^3P_0$ quark creation mechanism as
shown in Fig.\ref{fig1}(b). Then we give our numerical results and
discussion in Sec.4.
\\
\section{$\chi_{c0,2}\rightarrow\phi\phi$ in perturbative QCD}

 We first deal with the
$\chi_{cJ}\rightarrow\phi\phi (J=0,2)$ decays in pQCD scheme. The
annihilation of $\chicj$ can be described as $c\bar c$ quarks
annihilate into two gluons and then materializing into two
outgoing mesons as illustrated in Fig.1(a). In this work, we aim
to investigate the contribution from the color singlet $\chicj$
and only consider contributions from the leading order diagrams.
For the two-body decay, the partial decay width reads:

\begin{equation}
\label{eq1}
{ d\Gamma\over d\Omega_{cm}} = \frac{1}{2}\frac{p_{cm}}{32\pi^2M_{\chi_{cJ}}^2} \frac{1}{2J+1}
 \sum\limits_{\lambda,\lambda_1,\lambda_2} {\vert M_{fi}(\lambda,\lambda_1,\lambda_2) \vert }^2
\end{equation}
where $p_{cm}=\sqrt{M_{\chi_{cJ}}^2/4-M_{\phi}^2}$ is the momentum of outgoing mesons, and $J$ is the spin quantum
number of the mother meson, and
$M_{fi}(\lambda,\lambda_1,\lambda_2)$ is the transitional matrix element. As usual treatment, this matrix element
can be decomposed into hard-scattering part and the wave function of mesons involved,
and its  Mandelstam form reads:

\begin{equation}
\label{eq2}
M_{fi}(\lambda,\lambda_1,\lambda_2)=-iC\int
{\frac{d^4kd^4k_1d^4k_2}{(2\pi )^4(2\pi )^4(2\pi
)^4}\chi_{\chi_{cJ}}(k,P,\lambda)\hat T \chi^{*}_{\phi
1}(k_1,P_1,\lambda_1)\chi^{*}_{\phi 2}(k_2,P_2,\lambda_2)}
\end{equation}
where $C=2/{3\sqrt 3}$ is the color factor, and $P,P_1$ and $P_2$
are four momentum vectors for $\chicj$ and the two outgoing
mesons, respectively. $\hat T$ corresponds to the hard-scattering
part, which can be obtained with the standard Feynman rules. The
wave functions $\chi_i$ are taken as the Bethe-salpter(BS) form
\cite{7,8}. In their own center of mass (cm) frame, we have
\begin{eqnarray}
&&\chi_{\chi_{cJ}}(k,P,\lambda) = \frac{N}{2\sqrt{2}}\sum_{s_z,m}<
S,s_z ;L,m\vert J,\lambda>R_{\chi}(|\vec k|)Y_{1m}(\Omega_{\vec
k})\delta(P\cdot k/M_{\chi_{cJ}}) \nonumber\\ &&
\quad\quad\quad\quad\quad
\frac{1}{\sqrt{(E_{p_1}+m_c)(E_{p_2}+m_c)}}
(\sla{p}_1+m_c)(1+\sla{P}/M_{\chi_{cJ}})\sla{\varepsilon}(s_z)(-\sla{p}_2+m_c),\nonumber\\
&&\chi^{*}_{\phi_i}(k_i,P_i,\lambda_i)=\frac{N^{*}_i}{2\sqrt{2}}
R_{\phi}(|\vec k_i|)Y_{00}\delta(P_i\cdot k_i/M_{\phi})
\nonumber\\ &&
\quad\quad\quad\quad\quad\quad\frac{1}{\sqrt{(E_{p_{i1}}+m_c)(E_{p_{i2}}+m_c)}}
(-\sla{p}_{i2}+m_s)\sla{\varepsilon}^{*}(\lambda_{i})(1+\sla{P}_i/M_{\phi})(\sla{p}_{i1}+m_s).
\end{eqnarray}
with relations
$p_1=P/2+k,p_2=-P/2+k,p_{i1}=P_i/2+k_i,p_{i2}=-P_i/2+k_i$, and $R(q)$ is the radial wave function normalized as
$\int {R^2(q)q^2dq}=(2\pi)^3$.

For convenience, we change the matrix element in trace form. We have
\begin{eqnarray}
&&\chi_{\chi_{cJ}}(k,P,\lambda)T\chi^{*}_{\phi
1}(k_1,P_1,\lambda_1)\chi^{*}_{\phi 2}(k_2,P_2,\lambda_2)= \nonumber\\
&&~~~~~~~~~~~~~~~~~~~~~
g_s^4Tr[\chi_{\chi_{cJ}}(k,P,\lambda)\gamma^{\rho}\frac{i}{\sla{k}-\sla{k}_1+\sla{k}_2-m_c}\gamma
^{\mu}] Tr[\chi_{\phi 2}(k_2,P_2,\lambda_2) \gamma ^{\omega}
\chi_{\phi 1}(k_1,P_1,\lambda_1)\gamma ^{\nu}] \nonumber
\\&&~~~~~~~~~~~~~~~~~~~~~
G_{\mu\nu}(\frac{P}{2}-k_2+k_1)
G_{\rho\omega}(\frac{P}{2}+k_2-k_1)+(\chi_{\phi
2}(k_1,P_1,\lambda_1)\leftrightarrow\chi_{\phi1
}(k_2,P_2,\lambda_2)).
\end{eqnarray}
For simplicity, we take non-relativistic approximation to the
description of $c\bar c$ quarks due to their heavy quark masses.
For strange quarks, this approximate scheme is partly reasonable
as shown in literature\cite{majp}. Using these relations, $
N=i2\pi/\sqrt{m_c},N_i=i2\pi /\sqrt{m_s}, E_{p_1}=E_{p_2}=m_c$,
and $E_{p_{i1}}=E_{p_{i2}}=m_s$, we simplify the matrix elements
as:
\begin{eqnarray}
&&M_{fi}(\lambda,\lambda_1,\lambda_2)=g_s^4\frac{2}{3\sqrt{3}}NN^{*}_1N^{*}_2\phi_0^2\phi_1^{'}\frac{4}{\sqrt{2}m_c^5}
\varepsilon_{\mu\nu}(-2g_{\mu\nu}\varepsilon_1^{*}\cdot\varepsilon_2^{*}m_c^2+4\varepsilon_1^{*\mu}
\varepsilon_2^{*\nu}m_c^2+ \nonumber\\&&~~~~~~~~~~~~~~~~~~~~~
g_{\mu\nu}P\cdot\varepsilon_1^{*}P\cdot\varepsilon_2^{*}-\varepsilon_2^{*\nu}P_1^{\nu}P\cdot\epsilon_1^{*}+
\epsilon_1^{*\nu}P_1^{\nu}P\cdot\varepsilon_2^{*}-2P_1^{\mu}P_1^{\nu}\varepsilon_1^{*}\cdot\varepsilon_2^{*}
\nonumber\\&&~~~~~~~~~~~~~~~~~~~~~
+P_1^{\mu}\varepsilon_1^{*\nu}P\cdot\varepsilon_2^{*}-P_1^{\mu}\varepsilon_2^{*\nu}P\cdot\varepsilon_1^{*})
+(\varepsilon_1^{*}\leftrightarrow\varepsilon_2^{*},P_1\rightarrow
(P-P_1)),
\end{eqnarray}
with
\begin{eqnarray}
&&\varepsilon_{\mu\nu}(\lambda)\equiv\sum\limits_{m,s_z}<1,m;1,s_z|J,\lambda>\varepsilon^{\mu}(m)\varepsilon^{\nu}(\lambda_i);
 \varepsilon_i^{*}\equiv\varepsilon_i^{*}(\lambda_i), \nonumber\\&&
  \phi_0\equiv\int{\frac{d^4k_i}{(2\pi)^4}R_{\phi}(|\vec k_i|)Y_{00}\delta(P_i\cdot
  k_i/M_{\phi})}, \nonumber \\&&
\phi_1^{'}\varepsilon^{\mu}(m)\equiv\int{\frac{d^4k}{(2\pi)^4}R_{\chi}(|\vec
k|)Y_{1m}(\Omega_{\vec k})k^{\mu}\delta(P\cdot k/M_{\chi})}\nonumber.
\end{eqnarray}

With these relations, the decay width of $\ccjpp$ can be simply expressed as
\begin{equation}
\label{eq7}
\Gamma(\chi_{cJ}\rightarrow\phi\phi)=g_s^8\frac{(2\pi)^4\sqrt{m_c^2-4m_s^2}}{54m_s^{2}
m_c^{13}}\phi_0^4\phi_1^{'2}M(J),
\end{equation}
with
$$
M(0)=\frac{64}{3}(4m_c^4-16m_s^2m_c^2+48m_s^4)\pi,
$$
$$
M(2)=\frac{1}{5}\frac{128}{3}(13m_c^4+56m_s^2m_c^2+48m_s^4)\pi,
$$
where $m_c=M_{\chi_{cJ}}/2, m_s=M_\phi/2$, and $g_s$ is related to the strong coupling constant $\alpha_s$ by
$g_s^2=\alpha_s/4\pi$.
\\

\section{$\chi_{c0,2}\rightarrow\phi\phi$ based on $^3P_0$ mechanism}

In this part, we investigate the $^3P_0$ quark pair creation
scheme for these decays. The success of $^3P_0$ mechanism in light
hadron decays indicates that the creation of light quark pair
might play an important role in light hadron energy scale. Whether
$\tpz$ mechanism is also important to describe the creation of
strange quark pairs in $\chicj$ energy scale is still an open
question. For this purpose we present the $\tpz$ mechanism as
illustrated in Fig.1(b). We assumes that the initial $\chi_c$ at
first annihilate into an intermediate state $\chi_s$ and then it
propagates and transfers into $\phi\phi$ by  assuming a strange
quark-pair created from QCD vacuum. This process can be described
by a transition amplitude as
\begin{eqnarray}
&&M_{fi}(\lambda,\lambda_1,\lambda_2)=M_{\chi_{cJ}\rightarrow\chi_{sJ}}G_{\chi_{sJ}}M_{3P_0}(\lambda,\lambda_1,\lambda_2),
\end{eqnarray}
where $M_{\chi_{cJ}\rightarrow\chi_{sJ}},G_{\chi_{sJ}}$, and $M_{3P_0}$ correspond to the
sequential amplitudes as mentioned above. They are expressed as
\begin{eqnarray}
&&M_{\chi_{cJ}\rightarrow\chi_{sJ}}~~~~~~=~-ig_s^4C\int\frac{d^4k_1d^4k_2d^4k_3}{(2\pi)^4(2\pi)^4(2\pi)^4}
Tr[\chi_{\chi_{cJ}}\gamma^{\rho}S(\frac{P}{2}+k_1-k_2)\gamma_{\mu}]
G^{\mu\nu}(k_2)G^{\rho\omega}(P-k_2)
\nonumber\\&&~~~~~~~~~~~~~~~~~~~~~~~~~~~~~~~~~~~~~~~~~~~~~~~~~
Tr[\chi^{*}_{\chi_{sJ}}\gamma_{\nu}S(\frac{P}{2}+k_3-k_2)\gamma_{\omega}]+
(k_2\rightarrow(P-k_2),\nu\leftrightarrow\omega),\nonumber\\&&~~~~~~~~
G_{\chi_{sJ}}~~~~~=~\frac{1}{M^2_{\chi_c}-M^2_{\chi_{s}}+i\varepsilon},
\nonumber\\ \nonumber\\
&&M_{3P_0}(\lambda,\lambda_1,\lambda_2)=~
h(\lambda,\lambda_1,\lambda_2)\sqrt{64\pi^3M_{\chi_{sJ}}E_{\phi}^2} .
\end{eqnarray}
where $C=2/3$ is the color factor of this process, and $\chi_{sJ}$
is taken as BS wavefunction with $P^{'}=(M_{\chi_{sJ}},\vec 0)$.
Here, we introduce the factor
$\sqrt{64\pi^3M_{\chi_{cJ}}E_{\phi}^2}$  to match the phase space
as used in  \cite{Ackleh}, and $h(\lambda,\lambda_1,\lambda_2)$
corresponds the $\tpz$ amplitude \cite{Ackleh}, it reads

\begin{eqnarray}
&&h(\lambda,\lambda_1,\lambda_2)=-2g_I\int d\vec{k}\phi_{Am}(2\vec
k-2\vec B)\phi_B^{*}(2\vec k-\vec B)\phi_C^{*}(2\vec k-\vec
B)\frac{|\vec k|}{E_k} \sum
<\frac{1}{2},s_1;\frac{1}{2},s_2|1,\lambda_1>\nonumber
\\&&
~~~~~~~~~~~~~~~~~~~~~<\frac{1}{2},s_3;\frac{1}{2},s_4|1,\lambda_2>
<\frac{1}{2},s_2;\frac{1}{2},|1,s_z><1,s_z;1,m|J,\lambda>
<\frac{1}{2},s_1;\frac{1}{2},s_4|1,s_{z^{'}}>\nonumber
\\&& ~~~~~~~~~~~~~~~~~~~~~~<1,s_{z^{'}};1,m^{'}|0,0>Y_{1m^{'}}(\Omega_{\vec
k})\sqrt{8\pi}
\end{eqnarray}
with
\begin{eqnarray*}
&&\phi_{Am}(2\vec
k^{'})=R_{\chi_{sJ}}Y_{1m}(2\pi)^{\frac{3}{2}}=\frac{2^{3/2}|\vec
k^{'}|}{\sqrt{3}\pi^{1/4}\beta^{5/2}}e^{-k^{'2}/2\beta^2}Y_{1m}(\Omega_{\vec
k^{'}}) \nonumber\\ &&\phi_B(2\vec k)=\phi_C(2\vec
k)=\frac{1}{\pi^{3/4}\beta^{3/2}}e^{-k^{2}/2\beta^2},
\end{eqnarray*}
where $g_I$ is the strength of the quark pair creation.

For simplicity, we make use of the on-shell approximation to deal
with the two gluon propagators, namely,
\begin{eqnarray}
\frac{d^4k_2}{(2\pi)^4}G^{\mu\nu}(k_2)G^{\rho\omega}(P-k_2)
=2\frac{d^4k_2}{(2\pi)^4}g^{\mu\nu}g^{\rho\omega}(\pi)^2\delta(k_2^2)\theta(\omega_1)\delta[(P-k_2)^2]\theta(\omega_2)
=\frac{1}{64\pi^2}g^{\mu\nu}g^{\rho\omega}d\Omega_{\vec{k_2}}
\end{eqnarray}
where $k^0_2=|\vec k_2|=M_{\chi_{cJ}}/2$. As the previous
treatment on the pQCD scheme, we employ NR approximation for
calculation $M_{\chi_{cJ}\rightarrow\chi_{sJ}}$. Finally, we have
\begin{eqnarray}
&&M_{\chi_{c0}\rightarrow\chi_{s0}}=ig_s^4\frac{32\pi(2m_c^2+m_s^2)\phi_{1c}^{'}\phi_{1s}^{'}}{3(m_c^2+m_s^2)m_cm_s\sqrt{m_cm_s}}
\nonumber\\
&&M_{\chi_{c2}\rightarrow\chi_{s2}}=ig_s^4\frac{64\pi\phi_{1c}^{'}\phi_{1s}^{'}}{15m_cm_s\sqrt{m_cm_s}}
\end{eqnarray}

\section {Numerical Results and Discussion}

In the $\tpz$ scheme, there are three phenomenological parameters,
namely, the mass of intermediate state $M_{\chi s}$, the strength
of quark pair creation $g_I$ and the harmonic parameter of meson
wave-functions $\beta$. The mass $M^2_{\chi s}$ in propagator can
be neglected since it is much smaller than $M^2_{\chi c}$. The
strength $g_I$ is assumed to be the same as in light meson decays
\cite{Ackleh}, {\sl i.e.}, $g_I=0.5$GeV.  The harmonic oscillator
parameter $\beta$ takes the value $\beta_\phi=0.4$ GeV as used in
studies of the mass spectrum of light mesons. As for $\ccj$
states, the harmonic oscillator parameter is different from light
meson's, we take $\beta_{\chi_{c}}=0.47$ GeV \cite{pingrg}. In
pQCD scheme, for consistent consideration, the wave-functions of
mesons at origin are obtained from their harmonic oscillator form
associated with harmonic parameter as selected above, and the
strong coupling constant is taken as $\alpha_s$=0.3 \cite{Chiang}.
The numerical results for the $\ccj(J=0,2)\to\phi\phi$ decay
widths with the two mechanisms are tabulated in table 1.

\begin{table}[htbp]
\parbox{0.9\textwidth}{\caption{Numerical results for $\ccj(J=0,2)\to\phi\phi$ decay widths in
pQCD and $\tpz$ schemes as illustrated in Fig.1(a,b).}}
\vspace{0.5cm}
\begin{tabular}
{|p{101pt}|p{73pt}|p{85pt}|p{85pt}|p{80pt}|} \hline & pQCD& $^3P_0
(\beta=0.4$GeV)& $^3P_0 (\beta = 0.3$GeV)
&Expt. values \cite{PDG} \\
\hline $\Gamma (\chi _{c0} \to \phi \phi )$(kev)& 0.86& 60& 9.3&
10.1$\pm$ 7 \\
\hline $\Gamma (\chi _{c2} \to \phi \phi )$(kev)& 1.54& 1.1& 0.18&
$5.06^{+2.43}_{-2.03}$ \\
\hline $\frac{\Gamma(\chi _{c0}\to \phi \phi)}{\Gamma(\chi _{c2}
\to \phi \phi)}$& 0.56& 55 & 53 &
$2.0$ \\
\hline
\end{tabular}
\end{table}

In pQCD scheme, we find that the calculated decay width for
$\cxpp{2}$ is comparable with the experimental value,  while for
$\cxpp{0}$ the decay width is still suppressed as other
calculations for two pseudo-scalar final states. It is worth to
note that the ratio of their decay widths is independent on the
choice of parameter $\beta$. The calculated ratio
$\Gamma(\cxpp{0})/\Gamma(\cxpp{2})$ is about 0.56, much smaller
than the experimental value 2.0 \cite{PDG}. This indicates that
the $\cxpp{0}$ is highly suppressed in the conventional pQCD
scheme. In $\tpz$ scheme, we find that the calculated decay width
for the $\cxpp{0}$ decay is greatly enhanced. With adjustment of
parameters in reasonable ranges, we can reproduce the data very
well. For example, the results with $\beta=0.3$ GeV are listed in
the 4-th column of Table 1, which reproduce the data for
$\cxpp{0}$ quite well. In fact, due to larger energy scale here,
the strength $g_I$ could be smaller here than in the case of light
meson decays. The most important point for the $\tpz$ mechanism is
its very large ratio for $\Gamma(\cxpp{0})/\Gamma(\cxpp{2})$ of
about 54 which is rather independent of parameters involved. The
large ratio mainly comes from the much larger transition rate
between $c\bar c$ and $s\bar s$ through two on-shell gluons for
$0^{++}$ than for $2^{++}$ quarkonium, which is proportional to
multiplication of their corresponding $2\gamma$ decay widths. The
non-relativistic theoretical value for
$\Gamma(\chi_{c0}\to\gamma\gamma)/\Gamma(\chi_{c2}\to\gamma\gamma)$
is 3.75 while the experiment one is about 5. So the transition
rate for $0^{++}$ $c\bar c\to s\bar s$ is about 20 times larger
than the corresponding one for $2^{++}$. For the second step
$s\bar s\to\phi\phi$ through $\tpz$ quark-pair creation, the rate
is also larger for $0^{++}$ than for $2^{++}$. The resulted very
large $\Gamma(\cxpp{0})/\Gamma(\cxpp{2})$ ratio tell us that the
new $\tpz$ mechanism can give very large contribution to the
$\chi_{c0}$ decays, meanwhile gives little contribution to
$\chi_{c2}$ decays where the conventional pQCD mechanism may
dominate. Another enhancement factor for the $\tpz$ mechanism
comparing with the conventional pQCD mechanism is that the two
hard gluons involved can be on-mass-shell in the former but are
much off-mass-shell for the latter.

In summary, we have studied exclusive decays of $\chicj\to\phi\phi
(J=0,2)$ in both the conventional pQCD scheme and the new $\tpz$
mechanism. While the pQCD scheme can give a decay width for
$\cxpp{2}$ comparable with data, it gives a much smaller decay
width for the $\cxpp{0}$ decay. The new $\tpz$ quark pair creation
mechanism plays a dominant role for the $\cxpp{0}$ decay and is
expected to be also responsible for large rates for two
pseudo-scalar final states, while it has little influence to the
$\chi_{c2}$ decays.

\section*{Acknowledgements}
This work is partly supported by the National Nature Science
Foundation of China under grants Nos.  10225525,10055003,10447130
and by the Chinese Academy of Sciences under project No.
KJCX2-SW-N02.

\end{document}